\pdfoutput=1
\documentclass[acs,a4paper,preprint,unsortedaddress,onecolumn]{revtex4-1}
\usepackage{graphics}
\usepackage{color, colortbl}
\usepackage{amsmath,amssymb}
\usepackage{graphicx}
\usepackage{tabularx}

\newcommand{\vect}[1]{\textbf{\textit{#1}}}

\newcommand{\mo}[0]{\mathcal {O}}

\newcommand{\mh}[0]{\mathcal {H}}
\newcommand{\dist}[0]{\textrm {dist}}
\newcommand{\AT}[0]{\textrm{AT}}
\newcommand{\HY}[0]{\textrm{HY}}
\newcommand{\CG}[0]{\textrm{CG}}

\newcommand{\moleidxone}[0]{i}
\newcommand{\moleidxtwo}[0]{j}
\newcommand{\atomidxone}[0]{\alpha}

\newcommand{\thf}{{\textrm{th}}}
\newcommand{\rdf}{{\textrm{rdf}}}
\newcommand{\rep}{{\textrm{rep}}}
\newcommand{\dof}{{\textrm{DOF}}}
\newcommand{\exc}{{\textrm{extra}}}
\newcommand{\equi}{{\textrm{eq}}}

\newcommand{\confa}[0]{{\alpha_{\textrm{R}}}}
\newcommand{\confb}[0]{{\textrm{C}7_{\textrm{eq}}}}
\newcommand{\confc}[0]{{\alpha_{\textrm{L}}}}
\newcommand{\confd}[0]{{\textrm{C}7_{\textrm{ax}}}}

\begin{document}
\title{Adaptive Resolution Simulation in Equilibrium and Beyond}

\author{Han Wang}
\affiliation{CAEP Software Center for High Performance Numerical Simulation, Beijing, China}
\affiliation{Zuse Institute Berlin, Germany}
\email{wang\_han@iapcm.ac.cn}
\author{Animesh Agarwal}
\affiliation{Institute for Mathematics, Freie Universit\"at Berlin, Germany}

\begin{abstract} In this paper,
  we investigate the equilibrium statistical properties of both the force and potential interpolations of
  adaptive resolution simulation (AdResS) under the theoretical framework of
  grand-canonical like AdResS (GC-AdResS).
  The thermodynamic relations between the higher and lower resolutions are
  derived by considering the absence of fundamental conservation laws in mechanics for both branches of AdResS.
  % we discuss the necessary conditions for
  % accurate adaptive resolution simulation (AdResS) in equilibrium.
  % The statistical and thermodynamical properties of two AdResS
  % branches, i.e.~the force and the potential interpolations, are
  % discussed and compared in parallel.
  In order to investigate the applicability of AdResS method in studying the properties
  beyond the equilibrium, we demonstrate the accuracy of AdResS in computing the dynamical properties in
  two numerical examples: The velocity auto-correlation of pure water
  and the conformational relaxation of alanine dipeptide dissolved in water.
  Theoretical and technical open questions of the AdResS method are discussed in the end of the paper.
\end{abstract} %end of abstract
\maketitle

\section{Introduction}

Adaptive resolution simulation (AdResS)~\cite{praprotnik2005adaptive,praprotnik2006adaptive,praprotnik2007adaptive,praprotnik2008multiscale,poma2010classical,poblete2010coupling,praprotnik2011statistical,fritsch2012adaptive,bevc2013adaptive} is a concurrent multiscale simulation method for molecular system.
The terminology ``concurrent multiscale simulation'' means that
part of the system is simulated with higher resolution molecular models for the
purpose of accuracy, and \emph{in the meanwhile} the rest of the system is
simulated with lower resolution models for the purpose of saving computational cost.
AdResS is, therefore, suitable for the systems where different physical
phenomena is happening concurrently at different time and length scales. For example,
the structure of a protein and its solvation shell
should be resolved at the atomistic level, while far away form the protein,
only the hydrodynamic properties of the solvent are of interest, which can be studied with satisfactory
accuracy with very coarse-grained or even continuum
models~\cite{zavadlav2014adaptive,zavadlav2014adaptive1,delgado2009coupling}.
Similar  
techniques sharing the concurrent idea of multiscale simulation can be found, for example in Ref.~\cite{ensing2007energy,heyes2010thermodynamic,shi2006mixed,shen2014resolution,nielsen2010adaptive}.

The time evolution of a classical molecular systems
is modeled with Newtonian dynamics, and fundamental conservation laws
are satisfied, i.e.~the momentum and energy conservation.
However, in the AdResS system,
due to the spacial change of molecular resolution,
the two conservation laws cannot be satisfied in the same system~\cite{praprotnik2011comment,dellesite2007some}.
Traditionally, the AdResS scheme follows the \emph{force interpolation} approaches
that preserves the momentum conservation while breaks the energy
conservation.
Recently,
alternative approaches based on \emph{potential interpolation},
which preserve energy conservation while break the momentum conservation,
were designed~\cite{wang2013grand,potestio2013hamiltonian}.
The choice between the two AdResS branches therefore depends on the application. For example, in the
cases of studying hydrodynamics, the momentum conservation is preferred to the energy conservation.
However, if only the equilibrium properties are of interest,
the question should be asked on the ensemble that AdResS samples,
rather than directly on the mechanical conservation laws.
In this context,
it is worth noting that energy conservation is not a must for theoretical analyses on the equilibrium statistical properties,
although out of the energy conservation
one can define an auxiliary Hamiltonian~\cite{wang2013grand,agarwal2014chemical}
(not a physical one~\cite{dellesite2013multiscale}),
which reduces the mathematical difficulties.
In Sec.~\ref{sec:statistical} the force and potential interpolations are investigated 
under the same theoretical framework GC-AdResS (GC stands for ``Grand-Canonical like'')~\cite{wang2013grand} that raises necessary conditions
for a grand-canonical sampling with AdResS scheme.
In Sec.~\ref{sec:thermodynamic} the equilibrium thermodynamic relations between the resolutions are
provided for both branches of the AdResS.
From these relations, it is noticed that the absence of mechanical conservation laws 
modifies the way of thermodynamic equilibrium between the resolutions.

The GC-AdResS framework answers how to accurately compute the  equilibrium properties.
However, although the system is in equilibrium,
some properties, which are called dynamical properties, cannot be simply computed from the equilibrium ensemble averages.
% These properties are called dynamical properties.
A typical example is the velocity auto-correlation.
% that describs the infiniesimal response of the system
% to external perturbation.
Moreover, the time-scales of conformational transformations
of bio-macromolecules (e.g.~protein and DNA)
are, again, not equilibrium averages.
The investigation of these  dynamical properties are out of the scope of the GC-AdResS framework,
and to the best of our knowledge, there is no theoretical answers on how accurately  AdResS reproduces the 
dynamical properties of a system.
Therefore, it is of practical significance to  provide numerical
evidence on the performance of AdResS in computing the dynamical properties.
In Sec.~\ref{sec:dynamical}, we investigate the
velocity auto-correlation in an AdResS water system, and the
leading relaxation time-scales and corresponding conformational dynamics of
a model dipeptide system.
The importance of designing the thermostat in these simulations
is also discussed.
The paper is closed by a concluding section
that mainly discusses the open problems in AdResS from both theoretical
and practical perspectives.

\section{The hybrid resolution: bridging the molecular models of different resolutions}
\label{sec:design}

In most cases discussed in this work, we refer to higher resolution as the classical atomistic model, while
the lower resolution as the coarse-grained molecular model. However,
one should keep in mind that the ``atomistic'' and ``coarse-grained'' resolutions
may refer to more general molecular models, e.g.~the higher resolution can be
path-integral representation~\cite{poma2010classical}, and lower resolution can be continuous
hydrodynamic description~\cite{delgado2009coupling}.

\begin{figure}
  \centering
  \includegraphics[width=0.5\textwidth]{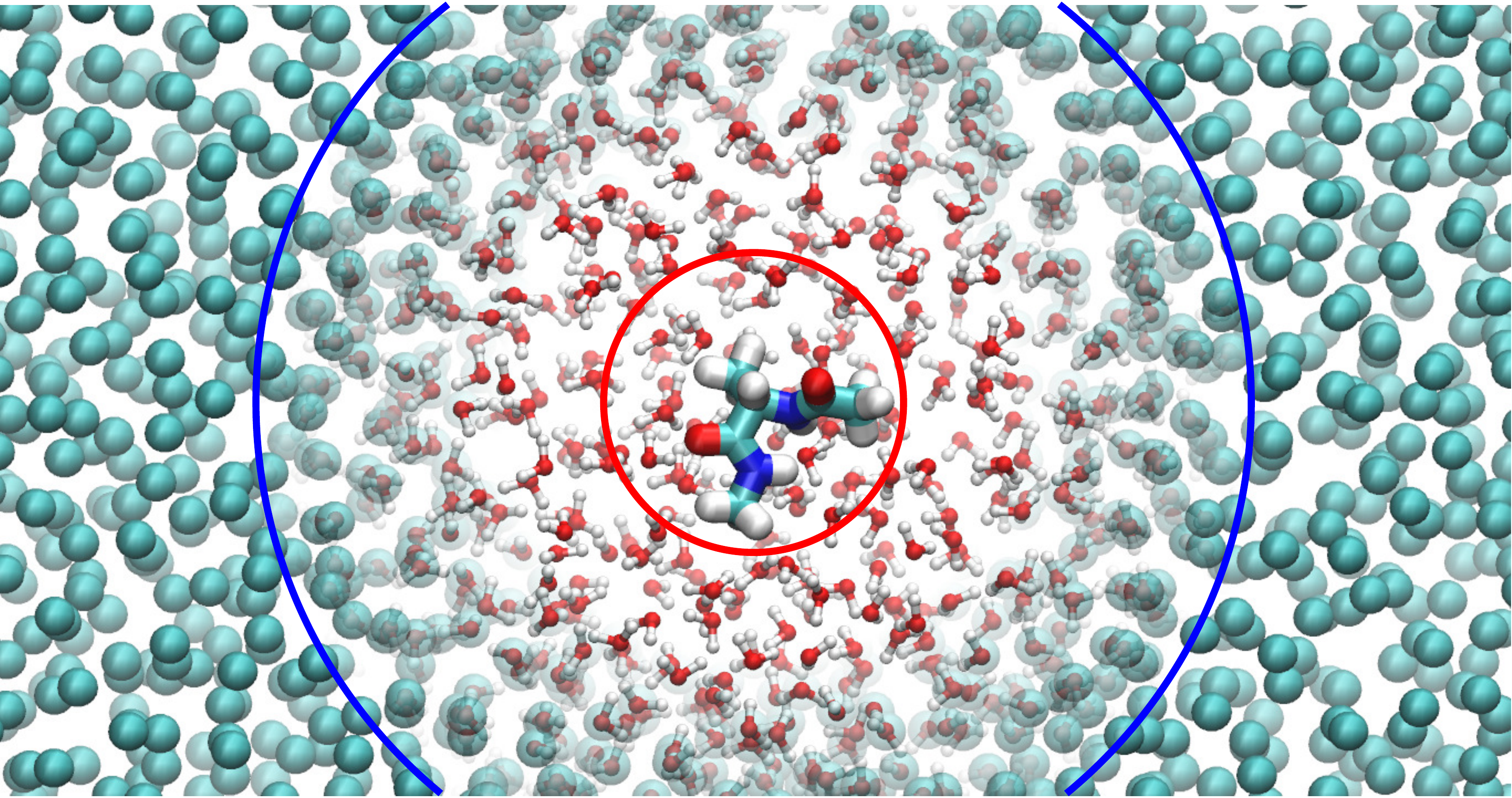}
  \caption{A schematic plot of an AdResS simulation. In side the red circle is the atomistic region. Between the red and blue circles is the hybrid region. Outside the blue circle is the coarse-grained region.}
  \label{fig:sys-region}
\end{figure}

In AdResS scheme, the simulation region $\Omega$ is 
decomposed without overlapping into three regions:
the atomistic region $\Omega_\AT$, the coarse-grained region $\Omega_\CG$, and
a hybrid region $\Omega_\HY$ bridging the former two regions.
That means the atomistic region is always connected to the coarse-grained region via a hybrid region.
The names of the atomistic and coarse-grained regions are self-explanatory, while
in the hybrid region each molecule has both the atomistic and coarse-grained resolutions.
A schematic plot of an AdResS simulation is presented in Fig.~\ref{fig:sys-region}.
The AdResS uses a scalar weighting function $w(\vect r),\ \vect r\in\Omega$ 
to denote the resolution of the molecule at position $\vect r$.
% In the atomistic region the value of the
% weighting function is usually set to 1, while in the coarse-grained
% region it is usually set to 0. In between, there is a bybrid region
% that is denoted by $\Omega_\HY$
% (or called transition region), where the molecules
% has both the atomistic and coarse-grained resolutions, and the weighting function
% goes smoothly from 1 to 0.
The definition of the weighting function
is not unique, one possible and perhaps the most popular choice is
\begin{equation}\label{eqn:old-w}
  w(\vect r) =
  \left\{
    \begin{alignedat}{3}
      &1 &\quad& \vect r \in \Omega_\AT\\
      &\cos^2\Big[\frac{\pi}{2 } \cdot \frac{\dist(\vect r, \Omega_\AT)}{d_\HY}\,\Big] && \vect r \in\Omega_\HY \\
      &0 &    & \vect r \in \Omega_\CG 
    \end{alignedat}
  \right.
\end{equation}
where $\dist(\vect r, \Omega_\AT)$ is the distance between the
position $\vect r$ and the atomistic region, which is defined
by $\dist(\vect r, \Omega_\AT) = \min_{\vect s\in\Omega_\AT} \vert
\vect r - \vect s \vert$.  $d_\HY$ is the thickness of the hybrid
region. We always assume the thickness of the hybrid region is
uniform, which indicates that the weighting function smoothly vanishes
at the boundary between the hybrid and the coarse-grained regions.
AdResS requires 
all interactions\footnote{The long-range electrostatic interactions
  are also treated by the cut-off method, e.g.~the Reaction-Field method.
} being treated
by the cut-off method, and the thickness of
the hybrid region being at least one cut-off radius (denoted by $r_c$).
Moreover, owing to the double-resolution of hybrid molecules,
an atomistic molecule only interacts with the 
atomistic resolution of a hybrid molecule,
while a coarse-grained molecule only interacts with the coarse-grained resolution
of a hybrid molecule.
The atomistic molecules are thus not directly interacting with the coarse-grained molecules.
The benefit of these settings is that
there is no extra work for modeling the cross-scale interactions.

The intermolecular interactions are
well defined in the atomistic and coarse-grained regions by the corresponding models.
To setup an AdResS simulation, one only needs to define the
interactions in the hybrid region. There are in general two
possibilities: force interpolation and the potential interpolation.
The force interpolation approach defines the hybrid force between two
molecules (indexed by $\moleidxone$ and $\moleidxtwo$) by a linear interpolation
between the atomistic (denoted by ${\vect F}_{\moleidxone\moleidxtwo}^{\AT}$) and coarse-grained (denoted by ${\vect F}_{\moleidxone\moleidxtwo}^{\CG}$) forces.
\begin{align}\label{eqn:f-f-interpol}
  {\vect F}_{\moleidxone \moleidxtwo}=w_\moleidxone w_\moleidxtwo{\vect F}_{\moleidxone\moleidxtwo}^{\AT}+(1-w_\moleidxone w_\moleidxtwo){\vect F}^{\CG}_{\moleidxone\moleidxtwo} 
\end{align}
where $w_\moleidxone = w(\vect r_\moleidxone)$ is the weighting function
measured at the center-of-mass (COM) position of molecule $\moleidxone$.
It should be noted that the AdResS  force interpolation is not conservative, i.e.~there does not
exist a potential so that the force interpolation is derived by taking the negative gradient of the potential,
unless the coarse-grained interaction is identical to the atomistic interaction~\cite{praprotnik2011comment,dellesite2007some}.
It is easy to show that the force interpolation satisfies the momentum conservation: Since
both the atomistic and coarse-grained forces are subject to the Newton's third law, i.e.~${\vect F}_{\moleidxone\moleidxtwo}^{\AT} = - {\vect F}_{\moleidxtwo\moleidxone}^{\AT}$
and ${\vect F}_{\moleidxone\moleidxtwo}^{\CG} = - {\vect F}_{\moleidxtwo\moleidxone}^{\CG}$. From the definition Eq.~\eqref{eqn:f-f-interpol} we have ${\vect F}_{\moleidxone\moleidxtwo} = - {\vect F}_{\moleidxtwo\moleidxone}$.

The potential interpolation approach defines the hybrid energy by
\begin{align}\label{eqn:v-v-interpol}
  {V}_{\moleidxone \moleidxtwo}=w_\moleidxone w_\moleidxtwo{V}_{\moleidxone\moleidxtwo}^{\AT}+(1-w_\moleidxone w_\moleidxtwo){V}^{\CG}_{\moleidxone\moleidxtwo}.
\end{align}
The intermolecular force is therefore calculated by taking the negative gradient on the potential ${\vect F}^V_{\moleidxone \moleidxtwo}= -\nabla_{\moleidxone}{V}_{\moleidxone \moleidxtwo}$,
which is explicitly written as
\begin{align}\label{eqn:f-v-interpol}
  {\vect F}^V_{\moleidxone \moleidxtwo}
  &=w_\moleidxone w_\moleidxtwo{\vect F}_{\moleidxone\moleidxtwo}^{\AT}+(1-w_\moleidxone w_\moleidxtwo){\vect F}^{\CG}_{\moleidxone\moleidxtwo}  - \nabla w_\moleidxone\cdot w_\moleidxtwo (V^\AT_{\moleidxone \moleidxtwo} - V^\CG_{\moleidxone \moleidxtwo})
\end{align}
Where $\nabla w_\moleidxone$ is the gradient of the weighting function measured at position $\vect r_\moleidxone$,
and the superscript ``$V$'' denotes that the force is defined
by the potential interpolation.
By  definition~\eqref{eqn:f-v-interpol}, the force of potential interpolation is conservative. 
The difference between the force and potential
interpolations lies in the last term  of Eq.~\eqref{eqn:f-v-interpol}, which is a force
acting along the direction of decreasing weighting function.
% This force is defined as \emph{the force of changing representation}.
It should be noted that
this force breaks the Newton's third law, therefore the force of potential interpolation
does not conserve the momentum.
We define the accumulative effect of this term by the \emph{force of changing representation}:
\begin{align}
  \vect F_\moleidxone^\rep = \sum_{\moleidxtwo}  \nabla w_\moleidxone\cdot w_\moleidxtwo (V^\AT_{\moleidxone \moleidxtwo} - V^\CG_{\moleidxone \moleidxtwo})
\end{align}

\begin{table}
  \centering
  \caption{A summary of the fundamental conservation laws in mechanics 
    satisfied by the normal molecular system and AdResS systems defined by force and potential interpolations.}
  \label{tab:conv-laws}
  \begin{tabular*}{0.8\textwidth}{@{\extracolsep{\fill}}lcc}\hline\hline
    &         Momentum Consv.     &       Energy Consv. \\
    Normal molecular System     &       {Yes}        &       {Yes}\\      
    AdResS force interpol.   &       {Yes}        &       {No}\\      
    AdResS potential interpol.  &     {No}  &       {Yes}\\\hline\hline
  \end{tabular*}
\end{table}
The fundamental conservation laws satisfied by both the force and potential interpolations
are summarized in Tab.~\ref{tab:conv-laws}. Unlike normal molecular systems, in which both
the momentum and energy are conserved, neither of the AdResS systems conserves
both laws. The breaks of these mechanics conservations have substantial influence on the thermodynamic
properties of the AdResS systems (see the discussions in Sec.~\ref{sec:thermodynamic}).
% The choice of the interpolation scheme depends on
% the practical requirement, for example, in hydrodynamic systems it is crucial to have the momentum
% conservation, then the force interpolation is prefered.
Regarding the equilibrium statistical properties, we will show later in Sec.~\ref{sec:statistical}
that both approaches approximately sample the grand-canonical ensemble.

When the intermolecular force is defined by either
Eq.~\eqref{eqn:f-f-interpol} or \eqref{eqn:f-v-interpol}, the total
force exerts on one molecule is the sum of all pairwise forces:
\begin{align}
  \vect F_\moleidxone = \sum_\moleidxtwo \vect F_{\moleidxone\moleidxtwo}, \quad \vect F^V_\moleidxone = \sum_\moleidxtwo \vect F^V_{\moleidxone\moleidxtwo}
\end{align}
The force on the atomistic degrees of freedom (DOFs) of a hybrid molecules is distributed 
by, 
\begin{align}
  \vect F_\atomidxone = \frac{m_\atomidxone}{M_\moleidxone}\vect F_\moleidxone, \quad   \vect F^V_\atomidxone = \frac{m_\atomidxone}{M_\moleidxone}\vect F^V_\moleidxone 
\end{align}
where  $\atomidxone$ is the indexes of atoms in molecule $\moleidxone$.
$m_\atomidxone$ is the mass of atom, and $M_\moleidxone$ is the total mass of molecule $\moleidxone$, i.e.~$M_\moleidxone = \sum_{\atomidxone\in\moleidxone}m_\atomidxone$.

Before discussing the statistical and thermodynamic properties of AdResS, a remark on the
choice of the weighting function should be added. 
In Ref.~\cite{wang2012adaptive} the authors introduced a modified weighing function with
a buffer region so that the atomistic region
interacts with the hybrid region only via the atomistic intermolecular interaction:
\begin{equation}\label{eqn:new-w}
  w(\vect r) =
  \left\{
    \begin{alignedat}{3}
      &1 &\quad& \vect r \in \Omega_\AT\\
      &1 && \vect r \in\Omega_\HY, \dist(\vect r, \Omega_\AT) < r_c \\
      &\cos^2\Big[\frac{\pi}{2 } \cdot \frac{\dist(\vect r, \Omega_\AT) - r_c}{d_{{\HY}} - r_c}\,\Big] && \vect r \in\Omega_\HY, \dist(\vect r, \Omega_\AT) \geq r_c\\
      &0 &    & \vect r \in \Omega_\CG 
    \end{alignedat}
  \right.
\end{equation}
The minimum thickness of the hybrid region is therefore $2r_c$.
It worth noting that the coarse-grained region does not require a buffer in the hybrid region, because
by definition the coarse-grained region interacts with the hybrid region only via the coarse-grained intermolecular interaction.
The new definition~\eqref{eqn:new-w} is crucial for the equilibrium statistical properties of the AdResS,
however, the extra cost is spent in the buffer $\{\vect r \vert \vect r \in\Omega_\HY, \dist(\vect r, \Omega_\AT) < r_c\}$,
which is treated in atomistic resolution. This cost is relatively small for systems with a large atomistic region.

% \recheck{Remark, add the remark of inversed weighting function}
% \vskip .2cm
% \noindent\textbf{Remark}: In the case where the high-resolution simulation
% is extremely expensive, e.g.~the path-integral AdResS~\cite{poma2010classical},
% the extra cost for simulating with buffer region is therefore not desirable.
% One possible solution is to assign the value of weighting function in an reversed way:
% it taks 0 in high-resolution region, while takes 1 in low-resolution region.
% With the reversed weighting, the buffer region is on the coarse-grained side, 

\section{Equilibrium statistical properties of adaptive resolution simulation}
\label{sec:statistical}

\begin{figure}
  \centering
  \includegraphics[width=0.8\textwidth]{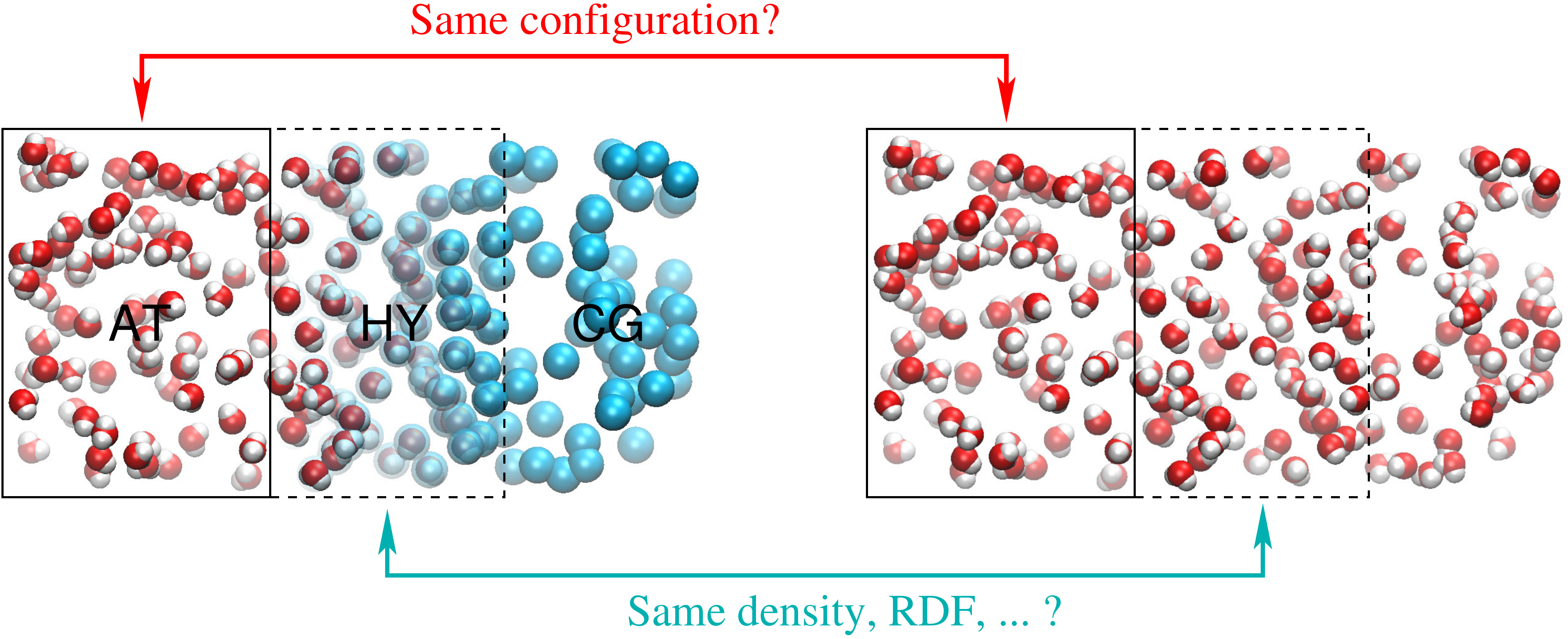}
  \caption{A schematic plot of the comparison between an AdResS system (left) and the full atomistic reference system (right).}
  \label{fig:compare}
\end{figure}

In order to consider the accuracy of the AdResS from statistical perspective, we always compare it with a full
atomistic reference system, which is of the same size as the AdResS system,
and contains the same number of molecules.
If the configuration of the atomistic region in the AdResS 
is same as the corresponding subregion in the atomistic reference, then
the AdResS  is of good accuracy (see the comparison indicated by the red arrow in Fig.~\ref{fig:compare}).
A well-known conclusion of the standard statistical mechanics is that
when the reference system approximates the thermodynamic limit, the subsystem is
subject to the grand-canonical ensemble. Therefore, it is natural to investigate
if the atomistic region of AdResS is also subject to the grand-canonical ensemble
in the same thermodynamic limit. The theoretical framework presented in this section works for
both the force and potential interpolations.

% the atomistic region is embeded into the AdResS system as if it were embeded in to an atomistic environment,
% and the AdResS  is of good accuracy (see the red arrow in Fig.~\ref{fig:compare}).
% Here the corresponding atomistic region in the reference sys
% \recheck{comments on the ``atomistic region'' of the reference system}

We denote the number of molecules, volume and temperature of the system by $N$, $V$ and $T$, respectively.
The atomistic reference system has exactly the same set of variables, and its equilibrium state  is
the \emph{desired equilibrium} of AdResS.
% The word ``desired'' means that
% the atomistic region of the AdResS system should reproduce the equilibrium properties corresponding subregion of the reference
% system.
The thermodynamic variables of the subregions are specified by the subscript,
for example, those of the atomistic region are 
$N_\AT$, $V_\AT$ and $T_\AT$. Those of the hybrid and coarse-grained regions
are denoted by adding subscripts ``$\HY$'' and ``$\CG$'', respectively.
Identities $N = N_\AT + N_\HY + N_\CG$ and $V = V_\AT + V_\HY + V_\CG$ obviously hold
in the AdResS system.
In equilibrium, the temperature is uniform across the system: $T = T_\AT = T_\HY = T_\CG$.
This is achieved in practice by coupling the whole system to a Langevin thermostat.
The pressure and chemical potential of the atomistic and coarse-grained regions
are denoted by $\{p_\AT, \mu_\AT\}$
and $\{p_\CG, \mu_\CG\}$, respectively.
The DOFs of molecules indexed $\moleidxone$ are $\vect x_\moleidxone = \{\vect r_\moleidxone, \vect p_\moleidxone\}$, where
$\vect r_\moleidxone$ denotes the generalized coordinates and $\vect p_\moleidxone$ denotes the corresponding momenta.
All the DOFs of the system are denoted by $\vect x = \{\vect x_1, \cdots, \vect x_N\}$. Without lost of
generality, we consider that molecules index by $\{1, \cdots, N_\AT\}$ are in the atomistic region, then $\{N_\AT+1, \cdots, N_\AT + N_\HY\}$ are in the hybrid region, and
the last $N_\CG$ molecules $\{N-N_\CG+1, \cdots, N\}$ are in the coarse-grained region. The corresponding DOFs are denoted by $\vect x_\AT$, $\vect x_\HY$ and $\vect x_\CG$.
It should be noted that for simplicity, we do not explicitly consider that the number of DOFs are different for an atomistic
and a coarse-grained molecule, and uniformly encode them by the vector $\vect x$.
% so it is acutally 
% \redc{The atomistic, hybrid and coarse-grained have different DOFs, a comment needed!}

% \begin{figure}
%   \centering
%   \includegraphics[width=0.3\textwidth]{figs/system.thermo/system.eps}
%   \caption{A schematic plot of the AdResS system in thermodynamic
%     equilibrium. The number of molecules, volume and temperature of
%     the atomistic and coarse-grained regions are denoted by $\{N_\AT,
%     V_\AT, T\}$ and $\{N_\AT, V_\AT, T\}$, respectively. The filter
%     allows free exchange of molecules between atomistic and
%     coarse-grained regions.}
%   \label{fig:system-thermo}
% \end{figure}

The thermodynamic limit is taken in the sense that both the atomistic
and the coarse-grained regions are infinitely large, and at the same
time, the atomistic region is much smaller than the coarse-grained
region. The hybrid region is much smaller than both the atomistic and
coarse-grained regions.  Therefore, the volumes of the
three regions satisfies $V_\CG\gg V_\AT\gg V_\HY$.  The coarse-grained
region can be treated as an infinitely large particle and energy
reservoir for the atomistic region, and the hybrid region is an
infinitely thin filter that changes molecular resolution when a
molecule passes by. In the
thermodynamic limit, if there were no change of resolution (or
considering the full atomistic reference system), it is obvious that
the subregion corresponding to the atomistic region samples the
grand-canonical ensemble.  The question regarding the accuracy of
AdResS can be asked by: How accurately the atomistic region samples
the grand-canonical ensemble in the thermodynamic limit? More
specifically, we want to prove the probability density of the atomistic region satisfies
\begin{align}\label{eqn:dist-0}
  p(\vect x_\AT, N_\AT) \approx \frac{1}{\mathcal Z} \exp\Big\{{\beta\mu^\ast_\AT N_\AT - \beta \mh^\AT(\vect x_\AT)} \Big\}
\end{align}
where $\mu^\ast_\AT$ is the chemical potential of reference system in
the desired equilibrium, $\mathcal Z$ is the partition function
normalizing the probability density, and $\mh^\AT$ is the \emph{atomistic}
Hamiltonian.
% defined in the the subregion of the reference system that
% corresponds to the atomistic region of the AdResS system.

Instead of directly proving Eq.~\eqref{eqn:dist-0}, Ref.~\cite{wang2013grand}
suggest investigating the following equivalent equations:
\begin{align}\label{eqn:dist-1-0}
  p(\vect x_\AT \vert N_\AT) &\approx \frac{1}{Z_{N_\AT}} \exp\Big\{{- \beta \mh^\AT(\vect x_\AT)}\Big\}  \\\label{eqn:dist-1-1}
  p(N_\AT) & \approx \frac{ Z_{N_\AT}}{\mathcal Z} \exp\Big\{{\beta\mu^\ast_\AT N_\AT}\Big\}
\end{align}
where $Z_{N_\AT}$ is the canonical partition function for an atomistic
system with $N_\AT$ molecules
\begin{align}
  Z_{N_\AT} = \int d\vect x_{\AT} \exp\Big\{{- \beta \mh^\AT(\vect x_\AT)}\Big\}.
\end{align}
The identity of the conditional probability holds: $ p(\vect x_\AT, N_\AT)  = p(\vect x_\AT \vert N_\AT) \, p(N_\AT) $.

\subsection{The accuracy of the configurational probability density}
The configurational probability density~\eqref{eqn:dist-1-0} is further split as
\begin{align}\label{eqn:conf-split}
  p(\vect x_\AT \vert N_\AT) =
  \sum_{N_\HY} \int {d}\vect x_\HY\,
  p(\vect x_\AT \vert N_\AT; \vect x_\HY, N_\HY)\cdot
  p(\vect x_\HY, N_\HY\vert N_\AT)  
\end{align}
The first probability density in the integral is the probability density
of atomistic DOFs conditioned on the number of atomistic molecules and all
hybrid DOFs.
It can be shown that if
(1) All interactions in the system are cut-offed;
(2) The atomistic region interacts with the hybrid region only in an atomistic way;
(3) The system is short-range correlated, and the correlation between the atomistic
and the coarse-grained regions is negligible, then the probability density $p(\vect x_\AT \vert N_\AT; \vect x_\HY, N_\HY)$
is approximated by
\begin{align}\label{eqn:dist-conf-0}
  p(\vect x_\AT \vert N_\AT; \vect x_\HY, N_\HY)
  \propto
  \exp\Big\{-\beta \mh^\AT(\vect x_\AT; \vect x_\HY, N_\HY) \Big\},
\end{align}
where 
\begin{align}
  \mh^\AT(\vect x_\AT; \vect x_\HY, N_\HY)
  =
  \sum_{i=1}^{N_\AT}\frac12 m_i\vect v_i^2 +
  \sum_{i,j=1}^{N_\AT} \frac12 V^\AT(\vect r_{ij}) +
  \sum_{i=1}^{N_\AT}\sum_{j=N_\AT+1}^{N_\AT + N_\HY} V^\AT(\vect r_{ij})
\end{align}
is the atomistic Hamiltonian with parameters $\vect x_\HY, N_\HY$. It
should be noted that the probability density~\eqref{eqn:dist-conf-0}
is identical to that of the reference system.
% with a subregion that
% corresponds to the hybrid region.
The probability density and local Hamiltonian
can be written down for the coarse-grained region analogically.

In general, the second probability density $p(\vect x_\HY, N_\HY\vert N_\AT)$ in Eq.~\eqref{eqn:conf-split} 
is not the same as the reference system.  For example, it has been
shown that the density distribution in the hybrid region deviates from
the desired one, even if the coarse-grained side is modeled to
reproduce the atomistic pressure~\cite{poblete2010coupling}.  However,
it is possible to raise necessary conditions
for the AdResS system, with which the accuracy can be improved
systematically (see the cyan arrow in Fig.~\ref{fig:compare}).
% improve the accuracy of the probability density.
The first necessary
condition is that the density profile of the hybrid region should be a
constant that is identical to the density of the atomistic reference
at desired equilibrium:
\begin{align}\label{eqn:necessary-1st}
  \rho_\HY(\vect r) = \rho_\AT^\ast = \frac NV
\end{align}
The second necessary condition is that the two body probability
density in the hybrid region is the same as the full atomistic
reference, In homogeneous and isotropic system, it is equivalent to
ask that the hybrid radial distribution function (RDF) should be the
same as that of the reference system:
\begin{align}\label{eqn:necessary-2nd}
  g_\HY(r) = g^\ast_\AT(r).
\end{align}
It is fully justified to systematically raise the necessary conditions
up to $m$-th multibody probability density, for example the third necessary condition would be
\begin{align}
  C^{(3)}_\HY = C^{(3)\ast}_\AT
\end{align}
where $C^{(3)}$ is the three-body correlation.
Since the system is homogeneous
and isotropic, the three-body probability density is equivalent to
three-body correlation function. 
% However, it is not convenient in
% numerics, because the correction to the $m$-th multibody probability density
% needs a $m$-body correction force in general, which is computationally
% too expensive.

In practice, the density in the hybrid region is corrected by the
thermodynamic force~\cite{fritsch2012adaptive}, which is applied on top of the AdResS intermolecular interactions:
\begin{align}\label{eqn:thf-f}
  \vect F_\moleidxone &= \sum_\moleidxtwo \vect F_{\moleidxone\moleidxtwo}  + \vect F_\moleidxone^\thf,  \\\label{eqn:thf-p}
  \vect F^V_\moleidxone& = \sum_\moleidxtwo \vect F^V_{\moleidxone\moleidxtwo} + \vect F_\moleidxone^{\thf,V},
\end{align}
where the thermodynamic forces is a one-body force defined over space,
i.e.~$\vect F^\thf_i = \vect F^\thf(\vect r_\moleidxone)$ and $\vect F^{\thf, V}_\moleidxone = \vect F^{\thf,V}(\vect r_\moleidxone)$,
for force and potential interpolations, respectively.
The thermodynamic force is applied only in the hybrid region, and is
calculated by the following iterative scheme:
\begin{align}\label{eqn:tfi}
  \vect F_{k+1}^{\thf(, V)} (\vect r) = \vect F_k^{\thf(, V)} (\vect r)-
  \frac{M}{\kappa(\rho^\ast_\AT)^2} \nabla\rho_k(\vect r)
\end{align}
where $k$ denotes the step of iteration and $\kappa$ is the
isothermal-compressibility. In equilibrium the density profile $\rho(\vect r)$ in the
atomistic and coarse-grained regions are constants, so the thermodynamic force
is updated only in the hybrid region. When the hybrid density is flat, the thermodynamic
force converges. An important consequence is that
\begin{align}
  \rho_\AT(\vect r) = \rho_\HY(\vect r) = \rho_\CG(\vect r) = \rho_\AT^\ast,
\end{align}
because the density profiles match at the atomistic-hybrid and
hybrid-coarse-grained boundaries.

By using a numerical example of SPC/E water~\cite{berendsen1987missing}
system, it has been shown in Ref.~\cite{wang2013grand} that when the
thermodynamic force is applied, the AdResS atomistic region reproduces (with satisfactory accuracy)
multi-body configurational probability densities of the reference system
up to the three-body correlation.
% In principle, higher order of mulit-body correlation can be investigated in
% a similar way.
A surprising numerical observation is that although
only the thermodynamic force is applied, the second order necessary condition~\eqref{eqn:necessary-2nd}
is automatically fulfilled, because the RDF is identical to the reference system in the buffer hybrid region
$\{\vect r \,:\, \vect r \in\Omega_\HY, \dist(\vect r, \Omega_\AT) < r_c\}$ that is interacting with
the atomistic region. The third necessary condition is not satisfied because there is deviation
in the three-body correlation in the buffer hybrid region.
It should be noted that there is no theoretical background why the second necessary
condition is satisfied by only using the thermodynamic force,
so one cannot expect the same benefit in other systems.

The second necessary condition~\eqref{eqn:necessary-2nd} is fulfilled by the
RDF correction $\vect F^\rdf$~\cite{wang2012adaptive}, which is a conservative two-body force added to the force interpolation~\eqref{eqn:f-f-interpol} as an extra term
that applies only in the hybrid region:
\begin{align}\label{eqn:rdf-corr-f}
  {\vect F}_{\moleidxone \moleidxtwo}
  =
  w_\moleidxone w_\moleidxtwo{\vect F}_{\moleidxone\moleidxtwo}^{\AT}
  +
  (1-w_\moleidxone w_\moleidxtwo){\vect F}^{\CG}_{\moleidxone\moleidxtwo}
  +
  w_\moleidxone w_\moleidxtwo(1-w_\moleidxone w_\moleidxtwo){\vect F}^{\rdf}_{\moleidxone\moleidxtwo}.
\end{align}
The RDF correction force is constructed from a RDF correction potential $V^\rdf$ by
\begin{align}
  \vect F_{\moleidxone\moleidxtwo}^\rdf = \vect F^\rdf(\vect r_{\moleidxone\moleidxtwo})
  = -\nabla_{\vect r} V^\rdf(r_{\moleidxone\moleidxtwo}),
\end{align}
which is calculated by the iterative Boltzmann inversion:
\begin{align}\label{eqn:ibi}
  V_{k+1}^\rdf(r) = V_k^\rdf(r) + k_BT \ln \Big[\frac{g_k(r)}{g^\ast_\AT(r)}\Big] 
\end{align}
where $k$ is the step of iteration. When the RDF of $k$th iteration is
identical to that of the full atomistic reference $g^\ast_\AT(r)$, the
iteration converges. It has been shown that the iterative scheme of
thermodynamic force~\eqref{eqn:tfi} coupled with the iterative Boltzmann inversion~\eqref{eqn:ibi}
is effective in correcting the density profile $g_\HY(\vect r)$ and
RDF $g_\HY(r)$ simultaneously~\cite{wang2012adaptive}.
For SPC/E water system when the RDF correction is applied, the third
necessary condition is automatically satisfied~\cite{wang2013grand}. Again, there is no
theoretical prove behind this phenomenon, and one cannot expect the same
benefit for other systems.

\noindent\textbf{Remark:} The RDF correction is developed only for the force interpolation scheme,
and a natural extension to the potential interpolation seems to be
\begin{align}
  {V}_{\moleidxone \moleidxtwo}
  =
  w_\moleidxone w_\moleidxtwo{V}_{\moleidxone\moleidxtwo}^{\AT}
  +
  (1-w_\moleidxone w_\moleidxtwo){V}^{\CG}_{\moleidxone\moleidxtwo}
  +
  w_\moleidxone w_\moleidxtwo(1-w_\moleidxone w_\moleidxtwo){V}^{\rdf}_{\moleidxone\moleidxtwo}.  
\end{align}
The effectiveness of this proposal has never been tested, and is still an open problem.

\subsection{The accuracy of the number probability}

The accuracy of number probability $p(N_\AT)$ is investigated by a
Taylor expansion w.r.t.~the relative size of the atomistic region to
the whole system, which vanishes in the thermodynamic limit. The sufficient condition for the first order accuracy
asks for a balance of the chemical potential~\cite{wang2013grand}:
\begin{align}\label{eqn:mu-eq}
  \mu_\CG - \mu_\AT = \omega_0,
\end{align}
where $\omega_0$ corresponds to the work from the filter on each
molecule that leaves the coarse-grained region and  enters the atomistic region. An important conclusion
form Ref.~\cite{wang2013grand} is that the relation~\eqref{eqn:mu-eq}
holds automatically when the thermodynamic force is applied so that the atomistic
and coarse-grained densities match. The conclusion is derived under
the assumption of thermodynamic limit and the decorrelation between the
atomistic and coarse-grained regions.

The second order accuracy is achieved if the isothermal-compressibility of the atomistic and coarse-grained resolutions match:
\begin{align}
  \kappa_\AT = \kappa_\CG
\end{align}
This is achieved by coarse-grained modeling, for example, the
structure based coarse-grained models that reproduce the atomistic
RDF~\cite{wang2009comparative}.

\subsection{The WCA potential as a generic energy and particle reservoir}

From the analysis on statistical mechanics, one surprising fact is that if the
requirement for the accuracy is not very high (only applying the
thermodynamic force is acceptable), then there is actually no
restriction on the coarse-grained model. One can even use ideal gas as
a coarse-grained model.  The computational difficulty lies in the
sudden switching-on of the atomistic interactions when a coarse-grained
molecule enters the hybrid region. If two molecules enter at
the same time and the same location, then the atomistic contribution
to the intermolecular interaction
% ~\eqref{eqn:f-f-interpol} or
% \eqref{eqn:f-v-interpol}
is infinity, which drives the simulation
unstable. This numerical difficulty can be avoided by using capped
atomistic interaction~\cite{praprotnik2005adaptive}, or by using a
gradually switching-on core-softened atomistic
interaction~\cite{heyes2010thermodynamic}.
In Ref.~\cite{wang2013grand}, the authors instead tested with
Weeks-Chandler-Andersen (WCA) potential~\cite{weeks1971role} as the
coarse-grained model, which is a short-ranged and pure repulsive
interaction.  For an SPC/E water system, the cut-off radius of WCA potential
can be 2.4 time smaller than the cut-off radius used in the atomistic
region. Since the computational cost of cut-off scheme is of order $\mo(r_c^3)$,
the pair interaction of WCA
is 19 times cheaper. Moreover, for each pair of molecules the atomistic model computes
10 pairwise interactions (9 electrostatic + 1 van der Waals), while the
WCA model computes only one. Therefore the WCA model in total costs only $1/190$ 
on force computation than the SPC/E atomistic model.
The WCA approach has been successfully used in calculating the
chemical potential of various complex fluids and mixtures~\cite{agarwal2014chemical}.

% Therefore
% certain manipulation in the hybrid region is need improve the accuracy
% of $p(\vect x_\HY, N_\HY\vert N_\AT)$. However, it is still not clear
% what is the sufficient conditions to achieve good accuracy in $p(\vect x_\HY, N_\HY\vert N_\AT)$.
% It is easy to raise necessary conditions: The marginal distributions
% of the hybrid cooridnates
% \begin{align}
%   p(\vect x_\HY) = \sum_{N_\HY = 0}^\infty \sum_{N_\AT = 0}^\infty p(\vect x_\HY, N_\HY\vert N_\AT) p(N_\AT)
% \end{align}
% should be the same as the reference system. Here we use the fact that
% the probability $ p(N_\AT)$ should be of good accuracy, which will be
% discussed later. This approach is validated because the discussion of
% accuracy of $ p(N_\AT)$ does not require a good accuracy of $ p(\vect
% x_\HY, N_\HY\vert N_\AT)$.  The $m$th marginal distribution of a probability density saying $p(\vect x_\HY)$ is defined
% by
% \begin{align}
%   p_\HY^{(m)}(\vect x_1, \cdots \vect x_m) = \int\cdots\int d\vect x_{m+1}\cdots d\vect x_{N_\HY}
%   p (\vect x_1, \cdots, \vect x_m, \vect x_{m+1}, \cdots \vect x_{N_\HY})
% \end{align}
% In a homogeneous system, the first marginal distribution is the
% density profile in the

% It is not desirable to explicitly write the 
      
% Eq.~\eqref{eqn:dist-1-0} and \eqref{eqn:dist-1-1} are equivalent to 

% In 
% A good accuracy means that the atomistic region is embeded into the AdResS
% To answer what is an accurate equilibrium AdResS simulation,
% To understand the statistical properties of an AdResS system, 

\section{Thermodynamic properties of adaptive resolution simulation}
\label{sec:thermodynamic}

In the reference system (full atomistic), if the equilibrium state is achieved, then
the temperature, pressure and chemical potential of different regions are identical.
In AdResS system, the global uniform
temperature distribution in equilibrium is equilibrated by the Langevin
thermostat.  However, extra terms appear in the pressure or chemical
potential relations when the momentum or energy conservation is absent
(see Tab.~\ref{tab:conv-laws}), respectively.

The chemical potential balance can be investigated from the fact that
when the uniform density distribution is guaranteed, the chemical potential
difference between the resolutions is compensated by $\omega_0$, which
is the work per molecule from the filter when a molecule leaves the
coarse-grained region and enters the atomistic region (see
relation~\eqref{eqn:mu-eq}).  Now the question turns out to be what
contributes to the work $\omega_0$.  For the potential interpolation,
the work of the filter is nothing but the work of the thermodynamic
force plus the extra kinetic energy corresponding to the extra DOFs
that a molecule obtains in higher resolution region:
\begin{align}\label{eqn:mu0-v}
  \omega_0 = \omega^{\thf,V} + \omega^\dof,
\end{align}
where $ \omega^{\thf,V}$ denotes the work of the thermodynamic force
of the potential interpolation.
This work is path independent if the thermodynamic force is defined as a function of the distance to the atomistic region, and applies along the gradient of the weighting function.
We always use the thermodynamic force that satisfies these conditions.
For the force interpolation, 
we denote the extra contribution due to the absence of the energy conservation
by $\omega^\exc$, then
\begin{align}\label{eqn:mu0-f}
  \omega_0 = \omega^{\thf} + \omega^\exc + \omega^\dof,
\end{align}
where $\omega^{\thf}$ is the work of the thermodynamic force of the force interpolation.
The magnitude of the extra term $\omega^\exc$ is unclear up to now.
% however, it is easy to computewith the help of the pressure relations.

For the force interpolation, the pressure of different resolutions
is related by~\cite{fritsch2012adaptive}
\begin{align}\label{eqn:p-eq-f}
  p_\CG - p_\AT = \rho_0 \:\omega^{\thf},
\end{align}
in which the work of the thermodynamic force again plays an important
role.  For the potential interpolation, the pressure relation can be
investigated by an imaginary infinitesimal volume increment of the atomistic
region, and the same amount of volume decrement of the coarse-grained
region. Following this idea, Ref.~\cite{agarwal2014chemical} proves that in
equilibrium the pressure is related by
\begin{align}\label{eqn:p-eq-v}
  p_\CG - p_\AT = \rho_0 (\omega^{\thf,V} - \omega^\rep)  
\end{align}
where $\omega^\rep$ is the work done by the average force of changing representation:
\begin{align}\label{eqn:work-rep}
  \omega^\rep = \int \langle \vect F^\rep(\vect r)\rangle_V\circ d\vect r
\end{align}
where ``$\circ$'' means that the integral should be understood as along the path.
The force of changing representation acts always along the gradient of the weighting
function, and the weighting function is a function of the distance to the atomistic region,
therefore, if the system is homogeneous in the hybrid region, then the integral~\eqref{eqn:work-rep}
is path independent. The ensemble average  can be estimated by sampling.
The force of changing representation stems from the last term of potential
interpolation~\eqref{eqn:f-v-interpol}, and contributes to the
pressure relation~\eqref{eqn:p-eq-v} because it does not satisfy the Newton's action-reaction
law~\cite{agarwal2014chemical}.

Using Eq.~\eqref{eqn:mu0-v} and \eqref{eqn:mu0-f} one has $\omega^{\thf,V} - \omega^\thf = \omega^\exc$,
while using \eqref{eqn:p-eq-f} and \eqref{eqn:p-eq-v} one has $\omega^{\thf,V} - \omega^\thf = \omega^\rep$.
Therefore, we have 
\begin{align}
  \omega^\exc = \omega^\rep
\end{align}
% which gives a explicit expression for the extra energy contribution
% due to the absence of energy conservation in the force
% interpolation.
It should be noted that the ensemble average in the work
of changing representation~\eqref{eqn:work-rep} is defined for the
potential interpolation, however, the estimate of the term
$\omega^\rep$ does not depend on the ensemble, at least up to the
``first order approximation''~\cite{agarwal2014chemical}, because the
hybrid region of the potential and force interpolation have the same
density. In Ref.~\cite{wang2013grand} an even stronger conclusion is shown numerically in the SPC/E water system: The
average force of changing representation estimated from potential
interpolation is pointwisly identical to that estimated from the force
interpolation. The thermodynamic relations discovered in
force and potential interpolation AdResS schemes are summarized
in Tab.~\ref{tab:thermodynamic}.

\begin{table}
  \centering
  \caption{A summary of the thermodynamic relations satisfied in force and potential interpolation AdResS.}
  \label{tab:thermodynamic}
  \begin{tabular*}{0.99\textwidth}{@{\extracolsep{\fill}}lll}\hline\hline
    &         Force interpol. AdResS     &       Potential interpol. AdResS \\
    Temperature    &   {$T_\AT = T_\CG$}                                                & {$T_\AT = T_\CG$}                                        \\
    Pressure       &   {$p_\CG - p_\AT = \rho_0\, \omega^\thf$}                          & {$p_\CG - p_\AT = \rho_0(\omega^{\thf,V} - \omega^\rep)$} \\
    Chemical pot.  &   {$\mu_\CG - \mu_\AT = (\omega^\thf + \omega^\dof +\omega^\rep)$}   & {$\mu_\CG - \mu_\AT = \omega^{\thf,V}+ \omega^\dof$}      \\\hline\hline
  \end{tabular*}
\end{table}

\section{Dynamical properties: adaptive resolution simulation beyond equilibrium}
\label{sec:dynamical}

% In Sec.~\ref{sec:statistical}, we investigated the accuracy of AdResS
% in sampling the equilibrium ensemble.
In principle, all quantities that
can be written down in terms of equilibrium ensemble averages should
be computed with high accuracy when using an AdResS simulation that approximately samples the grand-canonical ensemble.
However, under this framework one has little knowledge regarding the quantities
that are not ensemble averages.
For example, the non-equilibrium response of the system
with respect to an external perturbation, which is 
either computed by brute force non-equilibrium simulations~\cite{wang2014exploring}, or by
equilibrium time-correlation function via the Green-Kubo
relation~\cite{green1954markoff,kubo1957statistical} if the perturbation is small. The
time-correlation function, although measured in equilibrium
simulations, depends on how the dynamics
of the system is implemented.
The quantities of this type are called \emph{dynamical properties} of the system.
In principle, the Hamiltonian dynamics should be used to compute the dynamical properties,
however, due to the truncation error in computer simulations, it
is technically difficult to conserve the energy in very long MD simulations.
To this end thermostats are sometimes coupled to keep the system at desired thermodynamic
state.
It is worth noting that the thermostats are designed to sample equilibrium distributions, and 
disturb the Hamiltonian dynamics in an artificial way.
Therefore, when applying the thermostats
in computing the dynamical properties, the potential artifacts
should be checked carefully.
In the context of AdResS, we propose an alternative way of thermostating
the system, the local thermostat method~\cite{wang2014exploring},
which is proved to be free of the artifacts in computing the dynamical properties, in the following section.

% \footnote{In principle, 
%   whether the dynamical properties are disturbed by the thermostat should
%   be checked systematically with respect to the coupling parameters.}.
% Therefore, in the context of molecular dynamics simulation,

\subsection{Velocity auto-correlation of water}

\begin{figure}
  \centering
  \includegraphics[width=0.5\textwidth]{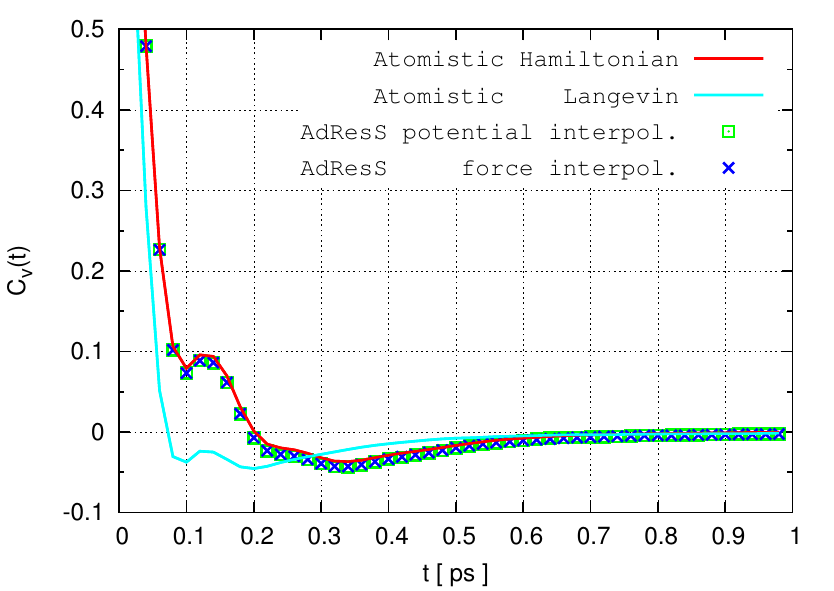}
  \caption{The oxygen velocity auto-correlation by atomistic
    simulation compared with AdResS simulation.  The solid cyan and
    red lines compares the atomistic simulation with and without
    Langevin thermostat, respectively.
    In the simulation of Hamiltonian dynamics, the energy drift is only 0.01\%.
    The square and cross points
    represent the potential and force interpolation AdResS with local
    thermostat, respectively.  The time-scales of the global and local Langevin
  thermostat are both $\tau_T = 0.1$~ps.}
  \label{fig:vac}
\end{figure}

In this section we take the normalized velocity auto-correlation function
for example:
\begin{align}
  C_v(t) = \frac{\langle \vect v(0)\cdot \vect v(t) \rangle}{\langle \vect v(0)\cdot \vect v(0) \rangle}
\end{align}
where the averages $\langle\cdot\rangle$ are still taken with respect
to the equilibrium ensemble, and the term
$\langle \vect v(0)\cdot \vect v(t)\rangle$ computes the correlation
between the velocity at time $0$ and $t$, which depends on the dynamics of the system. 
An illustrative example of the thermostating artifact is 
the oxygen velocity auto-correlation
in a TIP3P~\cite{jorgensen1983comparison} water system presented in Fig.~\ref{fig:vac}: If the system
is coupled to a Langevin thermostat, the result is qualitatively
different from that without any thermostat coupling, i.e.~Hamiltonian
dynamics. Although by using milder thermostats (increased time-scales)
will improve the result, it is in general difficult to know \emph{a priori}
the strength of the thermostat coupling.
The standard AdResS simulations, since the Langevin thermostat
is globally coupled to the system,  is not suitable for reproducing the dynamical
properties of the system.

The solution is to adopt the local thermostat method that was initially
developed for non-equilibrium MD simulations~\cite{wang2014exploring}.
This method divides the system into two regions.
The dynamical region, out of which the dynamical properties are calculated, is not coupled
to any thermostat, so the dynamics there is preserved as Hamiltonian dynamics.
The thermostated region is coupled to the Langevin thermostat in order to keep the system
at the desired thermodynamic state.
In the context of AdResS, we let the dynamical region identical to the atomistic region,
and
the hybrid and coarse-grained regions, which serve as particle and energy reservoir, coupled to the thermostat.
% The idea is to leave the atomistic region uncoupled to any thermostat, so the dynamics 
% is preserved as Hamiltonian dynamics.
% The region without thermostating is called the dynamics region. In this case the dynamical region
% is identical to the atomistic region.
% In the meanwhile, the hybrid and
% coarse-grained regions, which are serving as particle and energy reservior, are coupled to Langevin thermostat in order to keep the system
% at the desired  thermodynamic state. In this AdResS simulation, the dynamical properties
% should be measured only in the atomistic region.

This local thermostat method is validated by an AdResS simulation of pure TIP3P
water system.
The simulation is performed by the Gromacs~\cite{pronk2013gromacs} version 4.6.5.
The system contains 5000 water molecules.  The dimension of the periodic
simulation region is $14.7\times3.2\times3.2~\textrm{nm}^3$,
in which the resolution changes only in $x$-direction.
The atomistic region is of size $1.5\times3.2\times3.2~\textrm{nm}^3$, and the width of the hybrid region is $d_{\HY} = 2.85$~nm.
The rest of the system is coarse-grained region modeled by the WCA interaction.
The weighting function of form Eq.~\eqref{eqn:new-w} is used to couple different resolutions.
The Langevin thermostat coupled to the hybrid and coarse-grained regions is of time-scale $\tau_T = 0.1$~ps.
The thermodynamic forces of force and potential interpolations are iteratively computed by
Eq.~\eqref{eqn:tfi}.
The electrostatic interaction is treated by the Reaction-Field method~\cite{onsager1936electric,tironi1995generalized},
with dielectric constant $\varepsilon_{\textrm{rf}} = +\infty$, which conserves the energy, and
is proved to be a special case of the  Zero-multiple method~\cite{fukuda2011molecular,fukuda2013zero}.
The cut-off radius is 1.2~nm, and the van der Waals interaction is smoothed by the ``\texttt{switch}'' method provided by
Gromacs.

The Fig.~\ref{fig:vac} presents the velocity auto-correlation
functions computed by fully equilibrated MD trajectories of length 1~ns, with the
velocities recorded every 0.04~ps.  Here full equilibration means
uniform temperature and density distributions in the system.
In the Figure, both the force and potential interpolations are
in almost perfect agreement with the  atomistic simulation under Hamiltonian dynamics.
% presented along with the
% NVE atomistic simulation, and show almost perfect consistency.
This indicates that AdResS with local thermostat
is a promising method to investigate the dynamical properties
in the concurrent multiscale simulation.

% In the next subsection, we investigate the accuracy AdResS reprduces not only
% the equilibrium, but also the dynamical properties of a model
% poly-peptide system: Alanine dipeptide dissolved in water.

% \recheck{the difficulty of Langevin thermostat: destroy dynamics. Test local thermostating~\cite{wang2014exploring}
% }

\subsection{Alanine dipeptide dissolved in water: equilibrium and dynamical properties}

\begin{figure}
  \centering
  \includegraphics[width=0.99\textwidth]{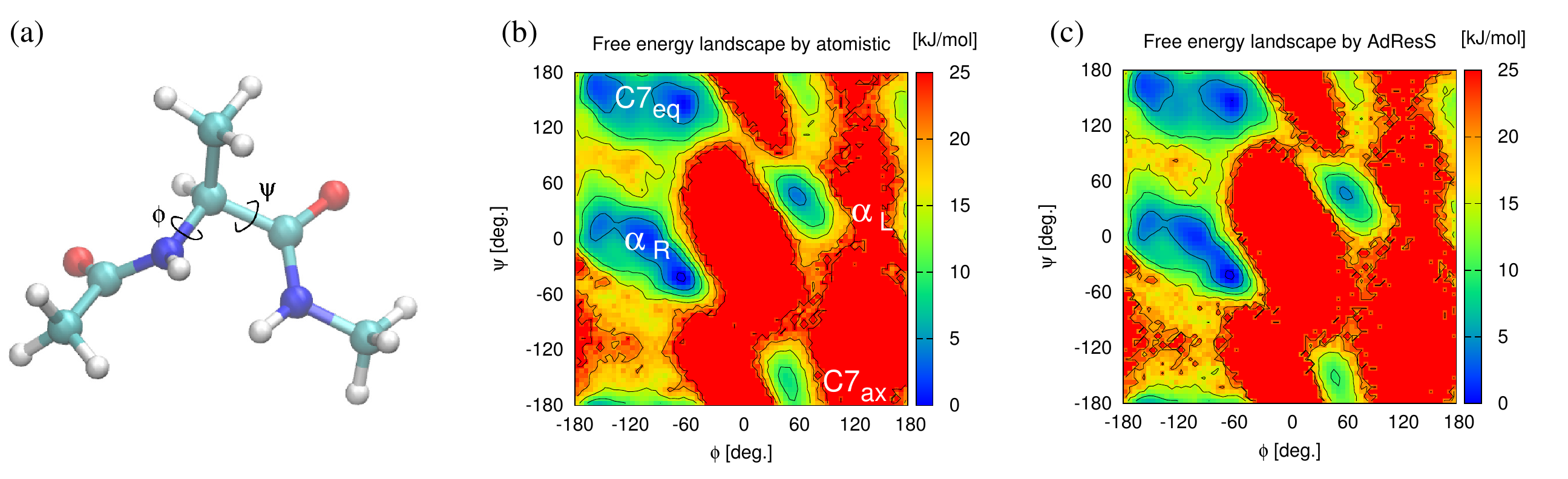}
  \caption{(a): A schematic plot of the alanine dipeptide molecule,
    and the two dihedral angles $\phi$ and $\psi$ that are used to
    represent the conformations of the molecule. (b) and (c): The free
    energy landscape plotted using variables $\phi$ and $\psi$.  (b)
    is for full atomistic reference simulation, while (c) is for
    AdResS. The metastable conformations coded by $\confa$, $\confb$, $\confc$ and $\confd$
    are noted on plot (b).}
  \label{fig:ala}
\end{figure}

The alanine dipeptide molecule (see Fig.~\ref{fig:ala}~(a)) is usually
used as a testing case for evaluating the new computational method in simulating 
biomolecular system (see e.g.~Ref.~\cite{apostolakis1999calculation,chodera2007automatic,kaminsky2007force,gfeller2007complex}).
We investigate a system composed of
7235 TIP3P water molecules, and
one alanine dipeptide described by the CHARMM27 force field~\cite{foloppe2000all} with
grid-based energy correction map (CMAP)~\cite{mackerell2004extending}. 
The dimension of the simulation region is $6.1\times6.1\times6.1~\textrm{nm}^3$, which is divided into
a spherical atomistic region of radius  $0.5$~nm centered at the $\alpha$-carbon and a  shell-shaped hybrid region of width 2.0~nm.
The rest of the system is treated coarse-grainly by the WCA interaction.
The dynamical region  is
also centered at the $\alpha$-carbon, and is of radius 1.2~nm.
It has been proved that the conformational dynamics of alanine dipeptide is not sensitive to the size of
the dynamical region when the radius is larger than 1.0~nm~\cite{wang2014exploring}.
In this case, part of the hybrid region is not thermostated, so
when using force interpolation, the extra energy production might not be effectively equilibriated.
To this end, we only investigate the potential interpolation. The corresponding
thermodynamic force is iteratively computed by
Eq.~\eqref{eqn:tfi}.
The electrostatic interaction is computed by the Reaction-Field method with dielectric constant $\varepsilon_{\textrm{rf}} = +\infty$.
Both the electrostatic and the van der Waals interactions (smoothed by ``\texttt{switch}'' method in Gromacs)
are cut-offed at 1.0~nm.

The system presents a very clear time-scale separation: There are fast
motions like bond and angle vibrations that happen in tens of femtoseconds, while
there are slow conformational transitions that happen in hundreds of picoseconds.
The dihedral angles $\phi$ and $\psi$ (see
Fig.~\ref{fig:ala}~(a)) are usually chosen as collective variables that  describe
the molecular conformation and its transitions.

The free energy is defined as the logarithm of the projected equilibrium probability density on the
$\phi$--$\psi$ space:
\begin{align}
  F(\phi,\psi) = -k_BT \ln p_\equi(\phi,\psi).
\end{align}
The free energy of the full atomistic reference system is showed in
Fig.~\ref{fig:ala}~(b), which clearly presents the metastability in the
system: several highly populated regions that correspond to metastable
conformations are separated by lowly populated regions that are
identified as transition regions.
The AdResS free energy (Fig.~\ref{fig:ala}~(c)) is in good
consistency with the full atomistic reference system, and
reproduces the metastable conformations and their populations
with satisfactory accuracy.
This is expected in equilibrium, because as discussed in
Sec.~\ref{sec:statistical}, the AdResS sampling approximates the
atomistic grand-canonical ensemble.

In practice, not only the equilibrium properties like the free energy are of interest,
but also the dynamical properties like the conformational transitions and the corresponding relaxation time-scales
are of highly biological importance.
% sometimes how the system relaxes to equilibrium is also of interest,
% e.g.~
% A direct analysis of the MD trajectories for these properties is not easy.
% Or equivalently
% at which time-scale the slow dynamics  happens and from which conformation to which it corresponds to.
When the system presents metastability,
one powerful tool for this purpose
is the Markov state model (MSM)~\cite{prinz2011probing,prinz2011markov,schuette2011markov}, which
approximates the original high-dimensional dynamics by a finite-state reduced
Markov process.
The accuracy analysis regarding  this approximation is well-established~\cite{sarich2010approximation,djurdjevac2012estimating}.
The MSM has found successfully applications in various fields, for example studying complex protein systems
(two of the hundreds of examples are Ref.~\cite{noe2009constructing,kohlhoff2014cloud}).

In a standard MSM approach, the full phase space is firstly decomposed into a non-overlapping union of finite number of states
(not necessarily identical to the metastable conformations),
then
the original time-dependent probability density define on the phase
space ($p_t(\vect x)$) is approximated by a temporal discrete probability $\vect p_{k\tau}$ defined on the finite-state space.
The step of the temporal discretization $\tau$ is called the lag-time.
The reduced
dynamics is assumed to be Markovian.
% although uaually it is not (or not perfectly Markovian),  and in practice it becomes one of the main sources of the error.
% We will not discuss the accuracy issue of the MSM in this paper,
% because it is not very relavant to the performance of the AdResS method.
% % It would be lengthy to discuss the topic here, and the interested readers are refereed to Ref.~\cite{}.
% % Due to the Markovianity of the reduced dynamics, the
Under the assumption of Markovianity, the original continuous dynamics is approximated by the Markov processes governed by
% probability at time $k\tau$ is given by
\begin{align}
\vect p^\top_{k\tau} = \vect p^\top_0 \cdot \vect T^k(\tau),
\end{align}
where $\vect T(\tau)$ is the transition matrix that
associates to the reduced dynamics of $\vect p^\top_{k\tau}$. The matrix element, for example $T_{ij}$
is defined as the probability of the system being in state
$j$ at $\tau$,  provided the system being in state $i$ at time 0.
% The ``$\cdot$'' denotes the matrix-vector multiplication.
In equilibrium, the transition matrix~$\vect T(\tau)$ does not depend on time when the state $i$ is investigated, and only the
lag-time $\tau$ matters.
% Due to the Markovianity, for any $k$, we have
% $p_{t+k\tau}(\vect x) = \mathcal Q^k(\tau)\circ p_t (\vect x)$.
If the transition matrix is irreducible and aperiodic, then according to Perron-Frobenius theorem, the maximum
eigenvalue is reached and single. Due to the stochasticity of the matrix, this eigenvalue is equal to 1.
If the dynamics
is further assumed to be reversible, then all eigenvalues of the transition matrix are
real valued.
The first $m$ largest eigenvalues are denoted by $1 = \lambda_1 > \lambda_2 \geq \lambda_3 \geq \cdots \geq \lambda_m$, and
the corresponding left eigenvectors are denoted by $\Phi_1, \Phi_2, \cdots, \Phi_m$.
The leading eigenvector is non-negative, and satisfies $\Phi^\top_1 = \Phi^\top_1\cdot\vect T(\tau)$.
With proper normalization (integrate to 1) the leading eigenvector is identical to the 
equilibrium probability density.
At infinitely long time, the probability density $\vect p_{k\tau}$ converges to the equilibrium probability.
The speed of the convergence is characterized by the leading relaxation time-scales that are
determined by the leading eigenvalues:  $t_i  = -{\tau}/{\log\lambda_i}, \ i=2, \cdots, m$.
Since the eigenvectors are orthogonal and the first eigenvector is non-negative, the reset
of the eigenvectors have a vanished sum over all elements: Their
value is positive at some states and negative at some other states.
The different signs in the eigenvectors denote the conformational transition corresponding to the
time-scale $t_i$.

% If the leading eigenvector is normalized to 1.
% The  reduced probability 

% The time evolution of the reduced probability density can be expaned by 
% \begin{align}\label{eqn:expan}
%   \vect p_{k\tau} = \Phi_1 + \sum_{i=2}^m e^{-\frac{k\tau}{t_i}} \,a_i\Phi_i + \textrm{fast decaying dynamics},
% \end{align}
% where $t_i$ is the time-scale that corresponds to the $i$th eigenvalue:
% \begin{align}\label{eqn:time-scale}
%   t_i  = -\frac{\tau}{\log\lambda_i}, \quad i=2, \cdots, m,
% \end{align}
% and $a_i$ are constant that are determined by the initial state. As can be seen from Eq.~\eqref{eqn:expan}, the
% time-dependent probability  converges to the equilibrium probability as time goes to infinity.
% The time-scale of the slow part of the relaxation is govenerned by the leanding eigenvalues through Eq.~\eqref{eqn:time-scale},
% while the conformational change corresponds to each time-scale is indicated by the eigenvector.
% We assume a gap in the spectrum between $\lambda_m$ and the rest of the spectrum $\lambda_{m+1}, \cdots$,
% therefore, the dynamics of the rest of the spectrum decays out very fast, and is not of particular interest.

% The spectral anaysis of the MSM will reveal the
% implied time-scales, and corresponding conformational changes.
% It has
% been proved that MSM is a good approximation to the original dynamics.

\begin{figure}
  \centering
  \includegraphics[width=0.5\textwidth]{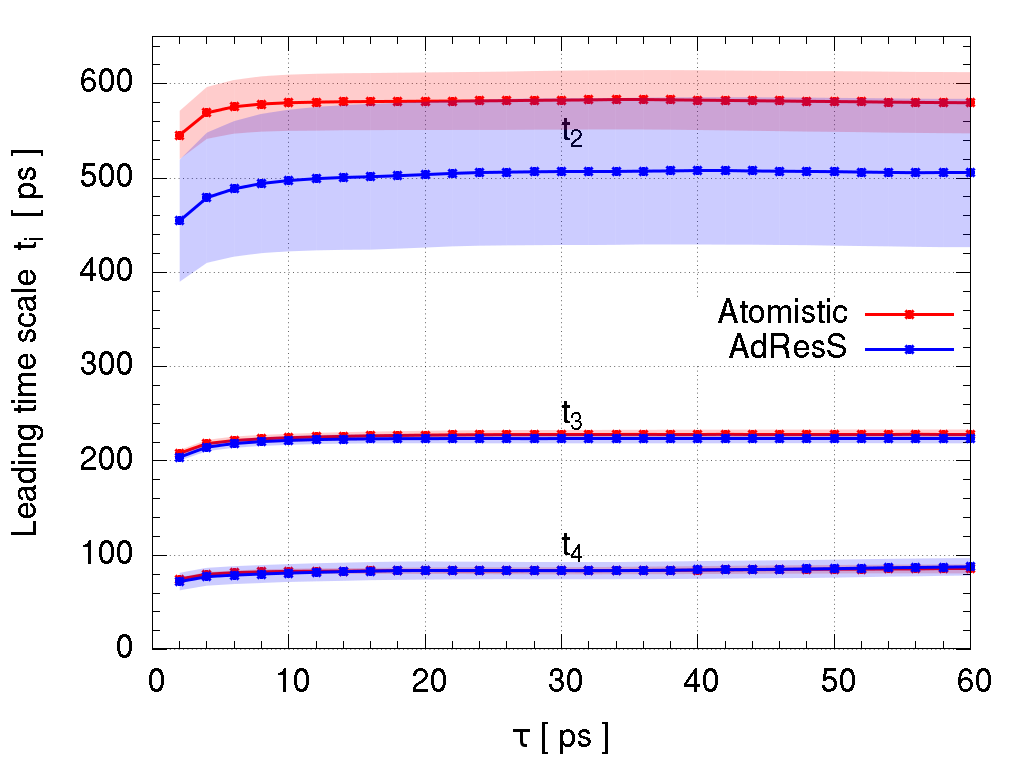}
  \caption{The non-trivial leading time-scale calculated from the
    Markov state modeling for the system of alanine dipeptide
    dissolved in water. In this figure, $t_2$ $t_3$ and $t_4$
    correspond to the first, second and third non-trivial leading
    time-scale, respectively. The horizontal axis $\tau$ is the
    lag-time used to build the Markov state model. For a sufficient
    large $\tau$ the values of the time-scales should be independent on
    the lag-time. The shadowing regions around the lines indicate the
    statistical uncertainty of the result. }
  \label{fig:ala-msm-t}
\end{figure}

\begin{figure}
  \centering
  \includegraphics[width=0.3\textwidth]{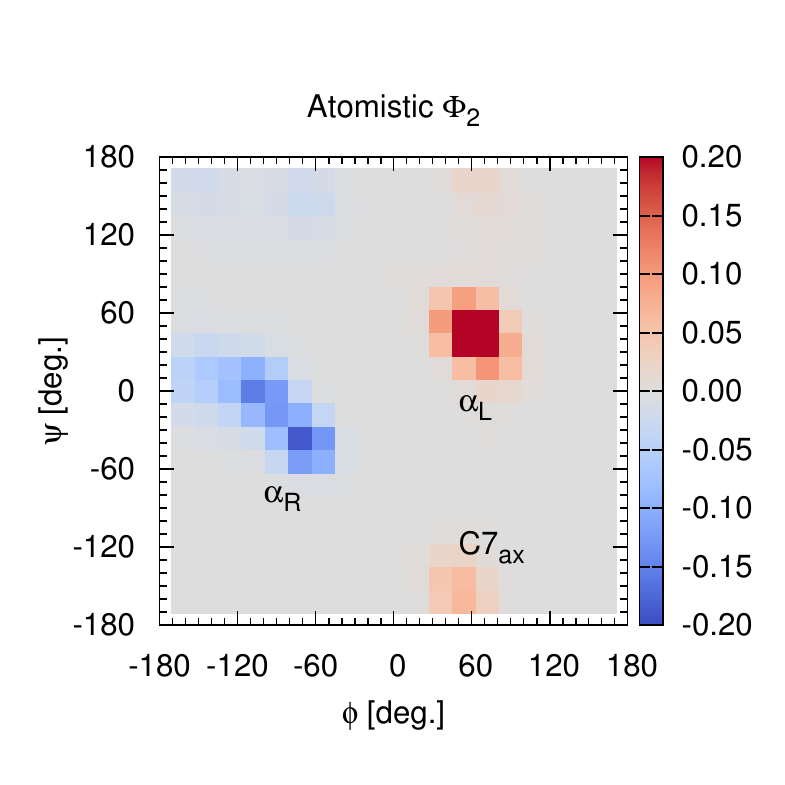}
  \includegraphics[width=0.3\textwidth]{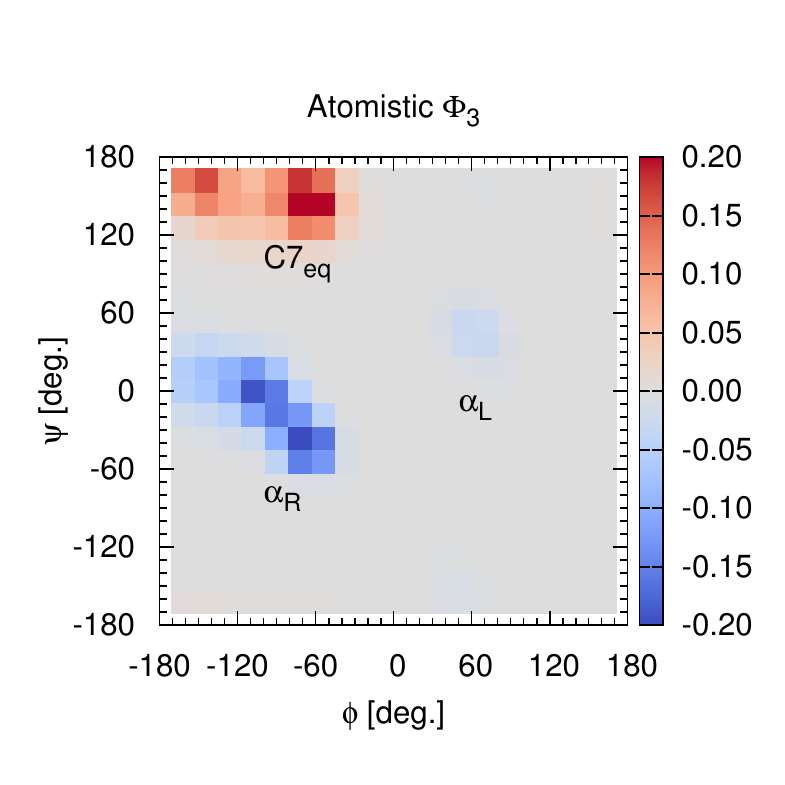}
  \includegraphics[width=0.3\textwidth]{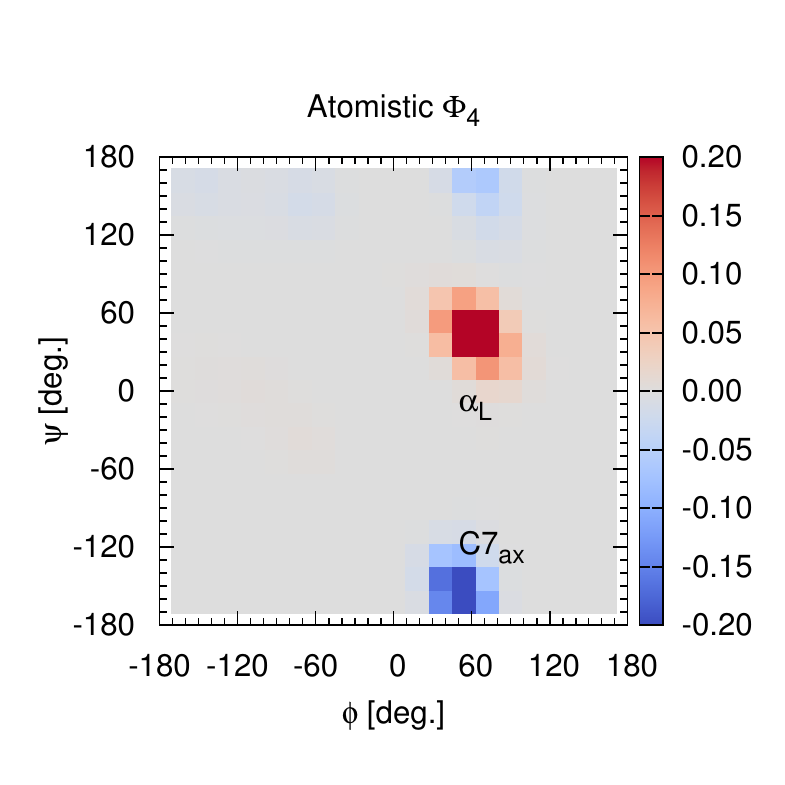}\\[-.4cm]
  \includegraphics[width=0.3\textwidth]{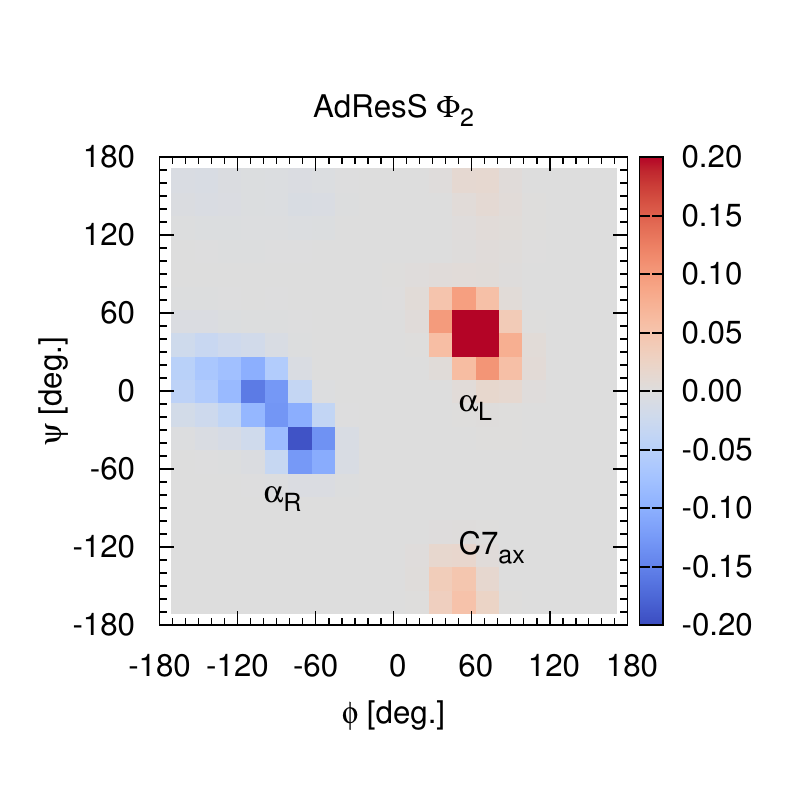}
  \includegraphics[width=0.3\textwidth]{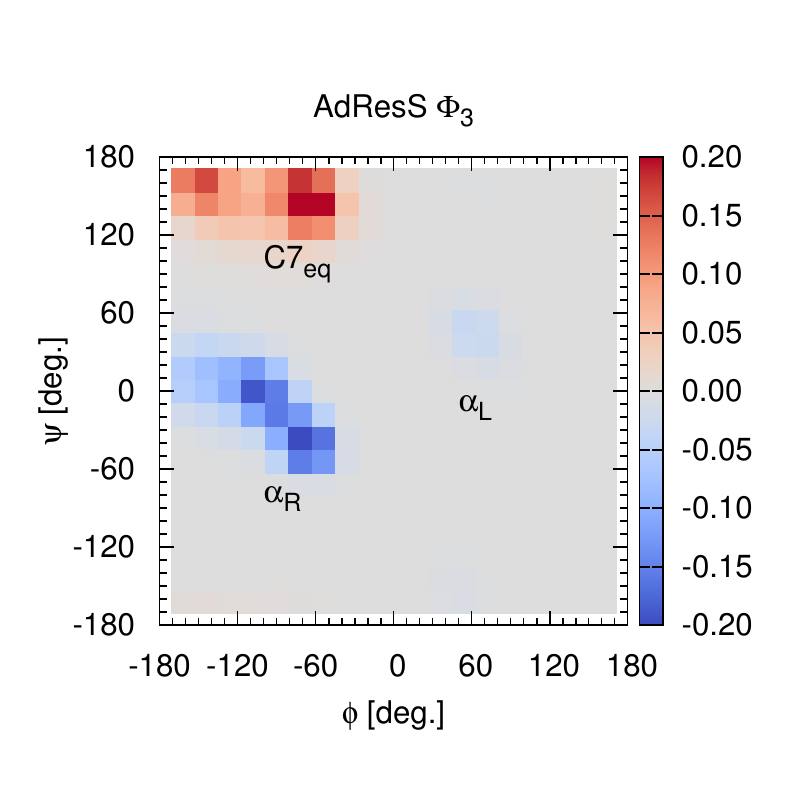}
  \includegraphics[width=0.3\textwidth]{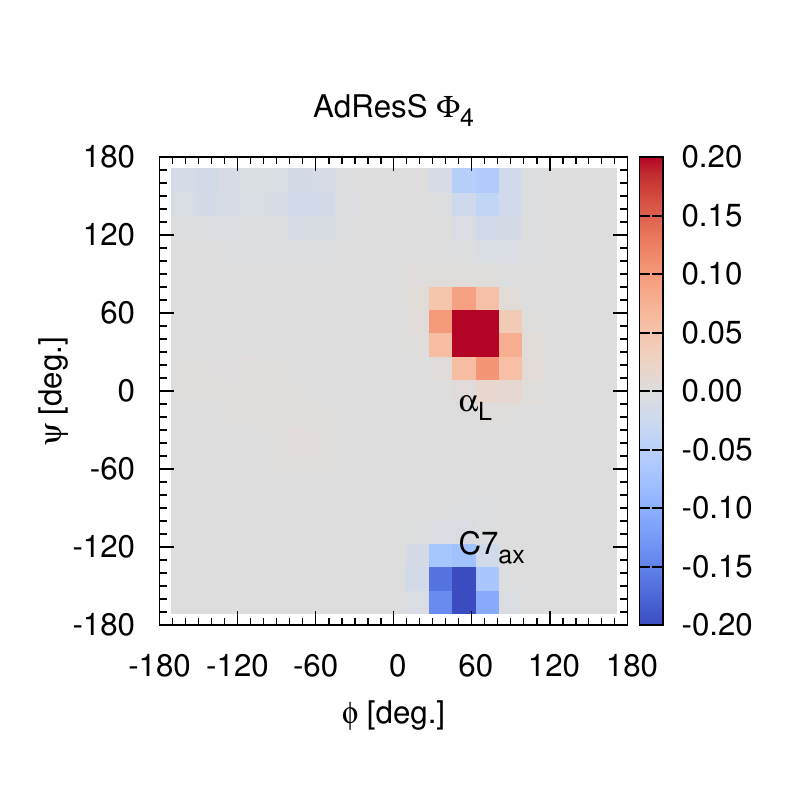}\\
  \caption{The eigenvectors corresponding to the first three leading time-scales computed using lag-time $20$~ps.
    The first row represents the eigenvectors of the full atomistic reference, while the second
    row represents those of the AdResS. From left to right $\Phi_2$, $\Phi_3$ and $\Phi_4$ that correspond to
    time-scales $t_2$, $t_3$ and $t_4$ are given (see also Fig.~\ref{fig:ala-msm-t}). The magnitude of the eigenvectors are
    denoted by the color. 
    The conformational
    changes indicated by the eigenvectors are noted on the plots.}
  \label{fig:ala-msm-v}
\end{figure}

In the example of alanine dipeptide dissolved in water, the states in the MSM
is defined by uniformly dividing the $\phi$-$\psi$ space in to $20\times 20$ bins.
Each bin presents a state in MSM, and the transition matrix is sampled by equilibrium
MD simulations with the local thermostating scheme~\cite{wang2014exploring}.
The MSM is built and analyzed by the software EMMA~\cite{senne2012emma}.
The resulting eigenvalues and eigenvectors of AdResS are compared with the full atomistic reference under the same local thermostat in
Fig.~\ref{fig:ala-msm-t} and \ref{fig:ala-msm-v}. The
eigenvalues should be independent for sufficiently large $\tau$ if the MSM is of good
accuracy, and from Fig.~\ref{fig:ala-msm-t} they are so for $\tau \geq 20$~ps.
Also from the Fig.~\ref{fig:ala-msm-t}, the full atomistic
reference and the AdResS eigenvalues are consistent within the statistical
uncertainty.
From Fig.~\ref{fig:ala-msm-v}, the conformational transitions correspond
to $t_2$, $t_3$ and $t_4$ are $\{\confa \} \leftrightarrow \{\confc,\confd\}$,
$\{\confa, \confc \} \leftrightarrow \{\confb\}$ and $\{\confc \} \leftrightarrow \{\confd\}$, respectively.
The eigenvectors and the indicated conformational transitions of the AdResS are also consistent with
those of the full atomistic reference. Although calculated from the
equilibrium MD simulations, it should be noted that the leading time-scales and the
corresponding eigenvectors are not the equilibrium properties, because
the transition probabilities (elements of the transition matrix) are actually
auto-correlation functions of the characteristic functions that are not equilibrium ensemble averages.

% A good reproduction of the equilibrium probability density does
% not necessary implies a reproduction of the auto-correlation functions.
% Although currently there is still no theoretical answer why AdResS works,
% the provided numercial example suggests that AdResS being promising in the applications beyond equilibrium,
% and  worth further theoretical studies.

\section{Concluding remarks and open questions}
\label{sec:conclusion}

In this paper, we investigated both the potential and force
interpolations under a uniform theoretical framework of GC-AdResS. The
necessary conditions for the grand-canonical like equilibrium
simulations are provided. However, it should be noticed that these
conditions are not sufficient, and the high accuracy in the  numerical simulations indicates
that these conditions tend to be too ``strong''.  Take the AdResS
water simulation for example, when only the thermodynamic force is
applied to impose the first necessary condition (density consistency),
the configuration of the atomistic region is almost identical to the
full atomistic reference up to the three-body
correlation\footnote{
  The three-body correlation is the highest order correlation function that we could reach.
  The four-body correlation is out of the capability of numerical
investigation, so it was not studied.}, and the second necessary condition (RDF consistency)
is automatically fulfilled.
% If the necessary conditions were not
% sufficient,
It might also be possible that the errors in the numerical simulations are too
small to be observed.  In this context, it would be helpful
to provide error estimates on the ensemble  sampled by AdResS with respect to the full atomistic
grand-canonical ensemble.
To the knowledge of the authors,
neither the sufficient condition nor the error estimate has ever been investigated.

If one were able to answer the questions regarding the accuracy of the equilibrium
ensemble of AdResS, the accuracy of quantities that are equilibrium ensemble
averages would be automatically answered. However, the dynamical properties, which are
not equilibrium ensemble averages, are out of the scope of the 
GC-AdResS framework. In order to numerically check the accuracy of AdResS
in computing the dynamical properties, we investigated the velocity auto-correlation
function of pure water and the leading relaxation time-scales of alanine dipeptide
dissolved in water.
The significant time-scale of the time-correlations ranges from 1~ps in the case of velocity auto-correlation
to almost 600~ps in the case of conformational relaxation of alanine dipeptide.
In both systems the AdResS reproduces the atomistic result
with a satisfactory agreement.
Although these positive numerically evidences cannot answer the open questions
on the reason behind the high accuracy, nor the error estimates of the dynamical properties,
they are strong evidence on the applicability of AdResS beyond equilibrium simulations,
and suggests a direction for theoretical investigations.

% These numerical result cannot answer the question
% of the theoretical reason behind the high accuracy, however, 

Currently the AdResS method has only been applied to model systems and small scale
applications. The advantage in the efficiency is not very obvious in these studies.
One important reason that prevents the AdResS method from
large scale applications is the software implementation~\cite{agarwal2014chemical}.
The software we used in the simulations is Gromacs\footnote{Versions higher than 4.6 support
AdResS.}.
Currently the force computation kernel for the hybrid region is written in C language~\cite{kernighan1988c}.
Comparing to the highly optimized assembly force computation kernels in Gromacs,
the AdResS kernel is not competitive. Moreover, the current AdResS implementation
keeps double resolution everywhere in the system\footnote{In the coarse-grained region,
  the atomistic DOFs are not used in the force computations, but they are kept and updated
  in every numerical integration step.
}.
Since in the coarse-grained region the force computation is very cheap, the majority
of the computational cost is actually spent on keeping and integrating the atomistic DOFs.
On straightforward solution is to remove the atomistic DOFs in the coarse-grained
region, which was actually proposed by the very original paper of AdResS~\cite{praprotnik2005adaptive}.
Owing to the fact that the coarse-grained interactions are usually ``softer''
than the atomistic interactions, one way of further boost the performance is
to integrate the coarse-grained molecules with larger time-steps.
If possible, the theoretical analyses on the
consistency and the corresponding
error estimates should be developed along the application of this adaptive time-step idea.
All these technical difficulties would hopefully be solved in the near future.

% \begin{thebibliography}{}
% % and use \bibitem to create references.
% \bibitem{RefJ}
% % Format for Journal Reference
% Author, Journal \textbf{Volume}, (year) page numbers
% % Format for books
% \bibitem{RefB}
% Author, \textit{Book title} (Publisher, place year) page numbers
% % etc
% \end{thebibliography}

\end{document}